\documentclass[12pt]{article}

\pdfoutput=1 

\usepackage{graphicx}
\usepackage{caption}
\usepackage{mathtools}
\usepackage{units}
\usepackage{bbold}
\usepackage{bm}

\newcommand{\dif}{\mathrm{d}}

\def\lapproxeq{\lower .7ex\hbox{$\;\stackrel{\textstyle <}{\sim}\;$}}
\def\gapproxeq{\lower .7ex\hbox{$\;\stackrel{\textstyle >}{\sim}\;$}}
\def\be{\begin{equation}}
\def\ee{\end{equation}}
\def\bea{\begin{eqnarray}}
\def\eea{\end{eqnarray}}

\def\GeV{\rm GeV}

\def\sh{\hat s}
\def\sh2{{\hat s}^2}

\def\J{\ensuremath{J/\psi}}
\def\Pb{\ensuremath{\mathrm{Pb}}}

\begin{document}

\begin{flushright}
DCPT/15/86 \\
FR-PHENO-2015-007 \\
IPPP/15/43  \\
LTP 1050 \\
MPP-2015-154 \\[2mm]
19th October 2015 \\
\end{flushright}

\vspace*{0.5cm}

\begin{center}
{\Large \bf Exclusive $J/\psi$ and $\Upsilon$ photoproduction}\\
\vspace*{0.5cm}
{\Large \bf and the low $x$ gluon}

\vspace*{1cm}

S.P. Jones$^{a, b}$, A.D. Martin$^c$,  M.G. Ryskin$^{c,d}$ and  T.Teubner$^{e}$\\

\vspace*{0.5cm}
$^a$ Albert-Ludwigs-Universit\"at Freiburg, Physikalisches Institut,
79104 Freiburg, Germany\\
$^b$ Max-Planck-Institute for Physics, F\"ohringer Ring 6, 80805
M\"unchen, Germany\\ 
$^c$ Institute for Particle Physics Phenomenology, University of
Durham, Durham\ \ DH1 3LE, U.K.\\
$^d$ Petersburg Nuclear Physics Institute, Gatchina, NRC Kurchatov
Institute, St.~Petersburg, 188300, Russia\\
$^e$ Department of Mathematical Sciences, University of Liverpool,
Liverpool\ \ L69 3BX, U.K.\\

\vspace*{1cm}

\begin{abstract}
We study exclusive vector meson photoproduction, $\gamma p \to V + p$
with $V=J/\psi$ or $\Upsilon$, at NLO in collinear factorisation, in
order to examine what may be learnt about the gluon distribution at
very low $x$. We examine the factorisation scale dependence of the
predictions. We argue that, using knowledge of the NLO corrections,
terms enhanced by a large $\ln(1/\xi)$ can be reabsorbed in the LO part
by a choice of the factorisation scale. (In these exclusive
processes $\xi$ takes the role of Bjorken-$x$.) Then, the scale dependence
coming from the remaining NLO contributions has no $\ln(1/\xi)$
enhancements. As a result, we find that predictions for the amplitude
of $\Upsilon$ production are stable to within about $\pm 15\%$.  This
will allow data for the exclusive process $p p \to p\Upsilon p$ at the
LHC, particularly from LHCb, to be included in global parton analyses
to constrain the gluon PDF down to $x\sim 10^{-5}$. Moreover, the
study of exclusive $J/\psi$ photoproduction indicates that the gluon
density found in the recent global PDF analyses is too small at low
$x$ and low scales. 
\end{abstract}
\vspace*{0.5cm}

\end{center}

\section{Introduction}
The present global PDF analyses (e.g.\ NNPDF3.0~\cite{Ball:2014uwa},
MMHT2014~\cite{Harland-Lang:2014zoa}, CT14~\cite{Dulat:2015mca}) find
that there is a large uncertainty in the low $x$ behaviour of the
gluon distribution. There is a lack of very low $x$ data, particularly
at low scales. Moreover, the gluon is determined at low $x$ mainly by
the DGLAP evolution of deep inelastic scattering, that is, not from a direct
measurement\footnote{In principle, measurements of $F_L$ would provide
  a direct determination, but the data are poor and do not reach low
  $x$. Moreover, the convergence of the perturbative series for $F_L$ is
  relatively poor.}, but rather from the derivative $\dif F_2/ \dif
\ln Q^2$. As a result, at low scales, the uncertainty on the gluon PDF
is large for $x \lapproxeq 10^{-3}$. On the other hand, the HERA data
on diffractive vector meson photoproduction
\cite{H1:1,H1:2,H1:3,H1:4,H1:5,ZEUS:1,ZEUS:2,ZEUS:3,ZEUS:4,ZEUS:5,ZEUS:6,ZEUS:7},
$\gamma p \to V + p$ and the LHC data on
the exclusive processes $pp\to p\,V\,p$~\cite{LHCb:1,LHCb:2,LHCb:3} where $V=J/\psi$ or $\Upsilon$, and $p\,\Pb \to p\,V\,\Pb$~\cite{TheALICE:2014dwa},
sample directly the gluon distribution down to $x \sim 10^{-5}$. For a review of exclusive vector-meson production at the LHC see, for example, Ref.~\cite{Baltz}.

However, $J/\psi$ and $\Upsilon$ data are not used in the global PDF
analyses.  One reason is that the corresponding cross sections are
described by non-diagonal (skewed) analogues of the PDFs, namely
generalised parton distributions (GPDs), due to the different masses
of the incoming photon and the outgoing $V$ meson~\cite{Ji}.  The 
GPDs can be related to the PDFs via the Shuvaev transform in the
low $\xi$ region~\cite{Shuv}.\footnote{For GPDs, $\xi$ plays the
  role of the PDF variable $x$ and is given by $\xi = (p^+ - p^{\prime +})/(p^+
+ p^{\prime +})$, where $p^+$ and $p^{\prime +}$ are the light-cone
plus-momenta of the in- and outgoing protons.}
In this work, following \cite{ShuvNockles}, we make the 
physically motivated assumption that the input distribution has no singularities in the right-half Mellin-$N$ plane
which implies that this relation holds at NLO to accuracy $\mathcal{O}(\xi)$. Another 
reason is the dependence of the theoretical predictions on the choice
of the factorisation scale, $\mu_F$. This problem has two independent
parts. One part has a technical nature and the other is more
physical. Let us discuss them in turn. First, the `technical' problem,
which is related to the convergence of the perturbative expansion at
low $\xi$ and low scales. A good illustration is that the NLO amplitude
for the exclusive high-energy $\gamma p \to J/\psi +p$
process~\cite{Ivan} was shown to yield a cross section which varies by
up to an order of magnitude for a reasonable variation of $\mu_F$. This
problem was emphasised recently by Wagner et al.~\cite{W-14}. The
strong scale dependence arises because in the DGLAP evolution of low
$\xi$ GPDs the probability of emitting a new gluon is strongly
enhanced by the large value of $\ln (1/\xi)$.
Indeed, the mean number of gluons in the interval $\Delta \ln \mu_F$
is~\cite{Dokshitzer:1978hw} 
\be
\langle n \rangle \;\simeq\; \frac{\alpha_sN_C}{\pi}\; \ln
(1/\xi)\;\Delta \ln \mu_F^2\,, 
\label{eq:n}
\ee
leading to a value of $\langle n \rangle$ up to about 8,  for the case
$\ln (1/\xi)\sim 8$ with the usual $\mu_F$ scale variation interval
from $\mu_F/2$ to $2\mu_F$.  In contrast, the NLO coefficient
function allows for the emission of only {\it one} gluon.
Therefore we cannot expect compensation between the contributions
coming from the GPD and the coefficient function as we vary the scale
$\mu_F$.  (At large $\xi$ the compensation is much more complete and
provides reasonable stability of the predictions to variations of the
scale $\mu_F$.) In Section~\ref{sec:OPT}, we use the NLO contribution
to fix a choice of the factorisation scale for the LO part of the
amplitude, which allows the $\ln(1/\xi)$ corrections to be resummed. 

The second or `physical' reason why there is a strong scale dependence
of $J/\psi$ and $\Upsilon$ photoproduction is due to `defects' in the
presently available PDF sets which are obtained in the DGLAP global
analyses of deep inelastic and related hard scattering data. That is, our studies indicate that there is a strong scale dependence in vector-meson photoproduction caused by the unexpectedly small LO contribution in comparison with the NLO correction. This occurs due to the very low input gluon density in the low $x$ domain obtained in the recent global PDF analyses.  We conclude that at the
input scale, the low $x$ gluon density is underestimated in comparison
with that for quarks. The crucial observation is that the number of
input gluons (which is parametrised {\it freely} in the global
analyses) is found to be much less than the number of such gluons
emitted by quarks --- or, to be more precise, gluons associated with
the quark (each quark carries a gluon field created by its colour
charge). With the present global partons the NLO sea quark
contribution to $V$ meson photoproduction grows with decreasing $x$
faster than the input gluons. As a result, at sufficiently high
energy, the quark component of the
NLO correction to $V$ photoproduction approximately cancels the main
LO contribution which is due only to the gluon density.  Therefore a
small variation of the scale in the NLO component leads to a strong
variation of the predicted cross section.

Recently the LHCb Collaboration have presented data for open charm and beauty production in the forward direction \cite{LHCbcc,LHCbbb}. These data sample approximately the same kinematic domain as exclusive vector meson production; actually the $x$ values are slightly larger, and the scale for beauty production is also larger.
It was shown in \cite{Blumlein,Gauld} that the data can be reproduced by NLO QCD using the present global PDFs. Note, however, that again there is a large sensitivity to the choice of factorization scale. Indeed, to reproduce the data one needs to take a scale approximately a factor of two larger than the natural choice $\mu_F=\sqrt{Q^2+m^2_Q}$. Since the optimal or appropriate scale for open heavy-quark production is not known at present, these data do not yet reliably determine the low $x$ gluon at low scales.  In the present paper we focus on exclusive vector-meson production. Moreover, our goal is not to present a new global PDF analysis or an explicit determination of the low $x$ gluon, but rather to study the possibility to reduce the scale dependence of the prediction, to study the qualitative structure of the NLO amplitude and to give hints of the consequences for the behaviour of the low $x$ gluon at low scales. 

In Section~\ref{sec:OPT} we consider the NLO contribution originating
from the light quark (singlet) GPD. We show that this
part of the NLO amplitude allows us to choose a factorisation scale
which sums the $\ln (1/\xi)$ enhanced
contributions inside the GPD. The remaining part of the NLO
contribution has no large $\ln (1/\xi)$ factors, and, as shown in
Section~\ref{sec:upsilon}, the compensation between the scale
dependence of the GPDs and of the coefficient functions for
$\Upsilon$ production makes the NLO result sufficiently stable 
for the data to be included in global parton analyses (just as
for the predictions of processes at larger $x$).  The study of the
$J/\psi$ photoproduction in Section~\ref{sec:J} suggests that the
gluon density obtained in the existing PDF analyses is too small at
low $x$ and low scale. 
In Section~\ref{sec:conclusions} we present our conclusions.

\section{NLO corrections and the choice of $\mu_F$ \label{sec:OPT}}
Our aim in this section is to show how knowledge of the NLO
contribution for $\gamma p \to V +p$ allows us to sum the
$\ln(1/\xi)$ enhanced contributions by a particular choice of $\mu_F$
in the LO part of the amplitude (and correspondingly in the NLO
coefficient function), which, in turn, reduces the factorisation scale
dependence of the predicted cross section. We outline the procedure in
Section~\ref{sec:LO}, but first we comment on the NLO formalism. 

\subsection{NLO formalism for $\gamma p \to V +p$  \label{sec:NLOgen}}
The NLO formalism for $\gamma p \to V p$ has been presented
in~\cite{Ivan}. However, before we can apply it to describe the high
energy $V$ photoproduction data, we must note one important
correction. 

The LO partonic amplitude, $\mathcal{A}_{g}^{(0)}(x/\xi)$,  for the
subprocess $\gamma +(gg) \to V$, can be calculated by considering the
fusion of a photon with a pair of on-shell gluons with zero transverse
momentum and physical, transverse, polarisations. If we use
dimensional regularisation with $D=4+2\epsilon$ (as in~\cite{Ivan}) to
regularize the ultraviolet (UV) and infrared (IR) divergences, then
the result is given, before dividing by the number of physical
transverse polarisations of the incoming gluon, by
\begin{equation} 
\mathcal{A}_{g}^{(0)}(x/\xi) = \alpha_{s}2(1+\epsilon)\,.
\label{eq:Ag0Step1}
\end{equation}
In $D$ dimensions, there are $D-2 = 2+2\epsilon$ possible transverse
directions. Since the entire calculation must be performed in $D$
dimensions, to properly average over the polarisation of the incoming
gluon, we must divide \eqref{eq:Ag0Step1} by this factor, and not
simply by 2.  Therefore for the LO partonic amplitude we obtain,
contrary to~\cite{Ivan} (prior to the first erratum), 
\begin{equation}
\mathcal{A}_{g}^{(0)}(x/\xi)=\alpha_{s}\,.
\label{eq:Ag0}
\end{equation}
The difference of a factor of $(1+\epsilon)$ does not alter the finite
LO result (as $\epsilon$ is taken to zero) but does lead to a
different NLO result due to the differing counter-terms. Indeed, the
extra factor $(1+\epsilon)$ presented in~\cite{Ivan} generates extra
terms $\sim\epsilon/\epsilon$ in the counter-terms, $\Delta$, giving
them a form 
\[
\Delta=\dots\left(\frac{1}{\hat{\epsilon}}+1+\ln\left(\frac{\mu_{F}^{2}}{\mu^{2}}\right)\right),
\]
where $1/\hat{\epsilon} = 1/\epsilon + \gamma_E - \ln(4\pi)$ and
$\gamma_E$ is the Euler-Mascheroni constant. However, with the LO
partonic amplitude \eqref{eq:Ag0} we instead  obtain counter-terms
containing 
\[
\Delta=\dots\left(\frac{1}{\hat{\epsilon}}+\ln\left(\frac{\mu_{F}^{2}}{\mu^{2}}\right)\right). 
\]
It is precisely this alteration to the counter-terms which leads us to
a different NLO finite result.\footnote{This correction was also
  mentioned in~\cite{CNocklesThesis, W-14} and discussed in more
  detail in~\cite{SJonesThesis}.} This result agrees with that
of~\cite{Ivan} with both errata. 

The details of our recalculation of the NLO result will be published
separately~\cite{GJT}. Here we simply state the final result for the
high-energy limit of the NLO part of the amplitude. In the high-energy
approximation, $W^{2}\gg M_V^2$ (where $W$ is the c.m. energy of
the incoming $\gamma p$ system and $M_V$ the mass of the vector
meson), the imaginary part of the amplitude dominates and the leading
contribution to the NLO correction comes from the region
$\xi\ll\left|x\right|\ll1$. The matrix element is given by 
\begin{eqnarray}
A^{(1)}(\xi,\mu_F) & \approx & -\mathrm{i}\pi \; C^{(0)}_g
\Bigg[\frac{\alpha_{s}(\mu_{R})N_{c}}{\pi}\ln\left(\frac{m^{2}}{\mu_{F}^{2}}\right)\intop_{\xi}^{1}\frac{\dif 
  x}{x}F_{g}(x,\xi,\mu_F) \nonumber \\  
 &  &
 +\;\;\frac{\alpha_{s}(\mu_{R})C_{F}}{\pi}\ln\left(\frac{m^{2}}{\mu_{F}^{2}}\right) \intop_{\xi}^{1}\dif
 x \left( F_{S}(x,\xi,\mu_F) - F_{S}(-x,\xi,\mu_F) \right) \Bigg]\,. 
\label{eq:he}
\end{eqnarray}
Here $m$ is the mass of the $c$ (for $\J$) or $b$ (for $\Upsilon$) quark and
\begin{align}
F_g(x, \xi, \mu) \; &= \; \sqrt{1-\xi^2} H_g(x, \xi, \mu), \\
F_S (x, \xi, \mu) \; &= \; \sqrt{1-\xi^2} H_S(x, \xi, \mu),
\end{align}
(for this unpolarised, forward process) with $H_g$, $H_S$ the gluon and quark singlet GPDs, respectively. The GPDs can be calculated from the diagonal PDFs, $q(x,\mu)$, $\bar{q}(x,\mu)$ and $g(x,\mu)$, via the Shuvaev transforms:
\begin{align}
H_S (x, \xi,\mu) \; &= \; \sum_{q=u,d,s} H_q(x,\xi,\mu) - H_q(-x,\xi,\mu), \\
\label{eq:shuvq}
H_q (x, \xi,\mu) \; &= \; 
\begin{dcases} 
\int_{-1}^1 \: {\rm d}x^\prime \left [ \frac{2}{\pi} \: {\rm Im} \: \int_0^1 
\: \frac{{\rm d}s}{y (s) \: \sqrt{1 - y(s) x^\prime}} \right ] \:
\frac{{\rm d}}{{\rm d}x^\prime} \left ( 
\frac{q (x^\prime,\mu)}{| x^\prime |} \right ), & x \ge 0\\
\int_{-1}^1 \: {\rm d}x^\prime \left [ \frac{2}{\pi} \: {\rm Im} \: \int_0^1 
\: \frac{{\rm d}s}{y (s) \: \sqrt{1 - y(s) x^\prime}} \right ] \:
\frac{{\rm d}}{{\rm d}x^\prime} \left ( 
\frac{\bar{q} (x^\prime,\mu)}{| x^\prime |} \right ), & x < 0
\end{dcases} \\
\label{eq:shuvg}
H_g (x, \xi,\mu) \; &= \; \int_{-1}^1 \: {\rm d}x^\prime \left [ \frac{2}{\pi} 
\: {\rm Im} \: \int_0^1 \: \frac{{\rm d}s (x + \xi (1 - 2s))}{y (s) \: \sqrt{1 - y (s) x^\prime}} 
\right ] \: \frac{{\rm d}}{{\rm d}x^\prime} \left ( \frac{g (x^\prime,\mu)}{| x^\prime |} \right ) \:,
\end{align}
where
\begin{equation}
\label{eq:a200}
y(s)=\frac{4s(1-s)}{x+\xi(1-2s)}\,,
\end{equation}
with an accuracy ${\cal O}(\xi)$, see Ref. \cite{ShuvNockles}.  Expressions (\ref{eq:shuvq}) and (\ref{eq:shuvg}) were used not only for the asymptotic NLO amplitude, (\ref{eq:he}), but also for the  full expression for the complete NLO amplitude used in the numerics presented below. The LO coefficient function, 
\begin{equation}
C_g^{(0)} =
\alpha_s(\mu_R) \frac{4\pi\sqrt{4\pi\alpha}e_{q}(e_{V}^{*}e_{\gamma})}{N_{c}\,\xi}\left(\frac{\langle
    O_{1}\rangle _{V}}{m^{3}}\right)^{1/2}, 
\label{eq:cLO}
\end{equation}
where $e_i$ are polarisation vectors and $\langle O_{1}\rangle _{V}$
is the NRQCD (non-relativistic QCD) matrix element of the $c{\bar c}
\to J/\psi$ (or $b{\bar b} \to \Upsilon$) transition.  In order to
have a small relativistic correction to $\langle O_{1}\rangle_{V}$, we
have to calculate the Feynman diagrams assuming that the
charm/bottom-quark line has mass $m=M_{V}/2$ where
$V=\J,\Upsilon$~\cite{Hood}. 

One can see directly from the high energy, leading $\ln(1/\xi)$,
limit, given in \eqref{eq:he}, that for the choice $\mu_{F}=m$ the asymptotic limit of the
quark NLO contribution vanishes. It is this observation that will
allow us (in Section~\ref{sec:NLO}) to claim that a suitable value for
the factorisation scale in leading logarithm terms is $\mu_F=m$. 

Before we show how the NLO contribution allows us to fix the scale
$\mu_F$ in the LO term, it is informative to first recall its
structure and introduce the notation in terms of the LO prediction. 

\subsection{General procedure \label{sec:LO}}
\begin{figure} [t]
\begin{center}
\includegraphics[width=0.4\textwidth]{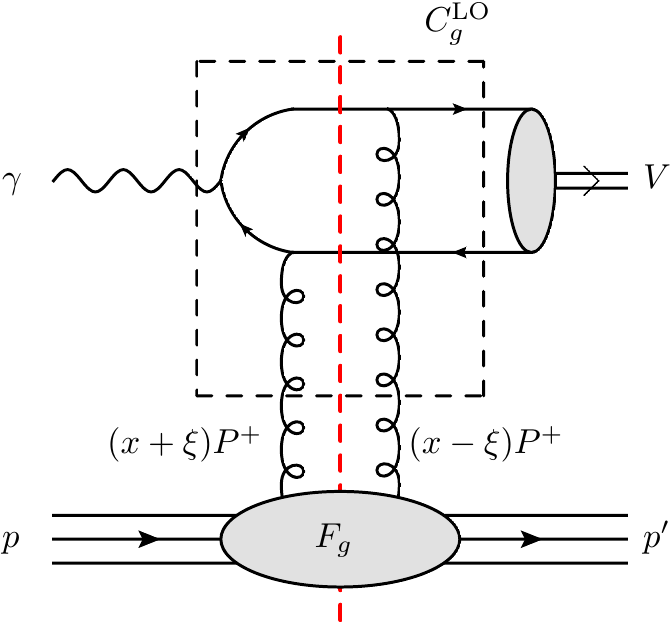}
\qquad
\includegraphics[width=0.4\textwidth]{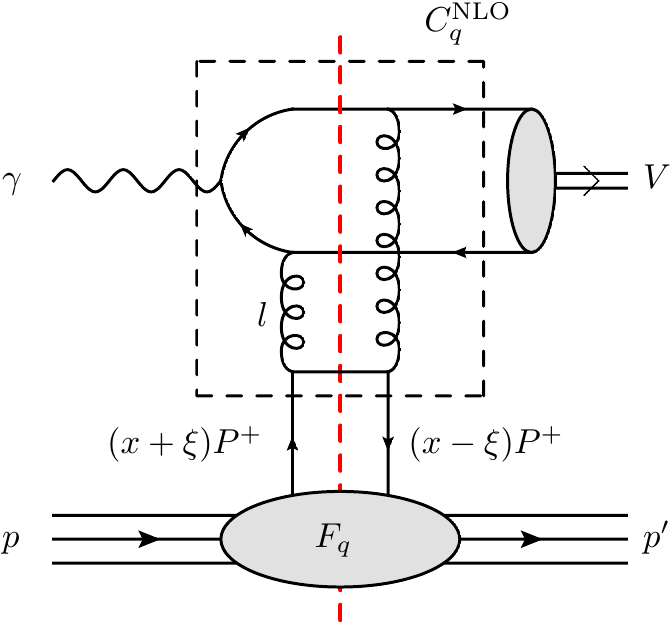}
\caption{\sf (a) The LO contribution to  $\gamma p \to V +p$, showing
  the convolution \eqref{eq:LO}. (b) The NLO quark contribution. For
  these graphs all permutations of the parton lines and coupling of
  the gluon lines to the heavy-quark pair are to be understood. Here
  $P\equiv (p+p^\prime)/2$ and $l$ is the loop momentum.
}
\label{fig:LO}
\end{center}
\end{figure}

We start with the LO contribution to $\gamma p \to V + p$. It is
sketched in Fig.~\ref{fig:LO}(a), and the amplitude is given by the
convolution 
\be
A^{(0)}(\xi,\mu_F)\;= \int^1_{-1}\frac{dx}{x}C^{(0)}_a(x/\xi)
F_a(x,\xi,\mu_F)~\equiv ~  C^{(0)}_a \otimes F_a(\mu_F)\,, 
\label{eq:LO}
\ee
where the sum over $a=q,g$ is understood and $C^{(0)}_q=0$ for this
process. The coefficient function $C^{(0)}_g$ is calculated using the
non-relativistic vector meson wave function. In general, the
relativistic corrections are not small for the $J/\psi$ case. These
corrections should be considered together with the three parton
($c\bar c+g$) component of the wave function; that is, accounting for
the r\^{o}le of gluons which provide the interaction between charm
quarks. For this process, it was shown by Hoodbhoy~\cite{Hood} that the
consistent treatment of relativistic corrections may, to good
accuracy, be effectively accounted for by choosing, in the
non-relativistic formula, the charm quark mass
$m_c=M_{J/\psi}/2$.\footnote{Strictly speaking, Hoodbhoy considered the electroproduction limit $Q^2\gg  M^2_{J/\psi}$. This limit simplifies the calculation, but inspection of the proof indicates that the statement about a small relativistic correction should be reliable below this limit as well.} After this, the remaining part of the correction
is quite small (a few percent only). So, we may use the
non-relativistic $J/\psi$ wave function with $m_c=M_{J/\psi}/2$, to
obtain a result with good accuracy. 

We return to discuss $V=J/\psi$ or $\Upsilon$. At LO the coefficient
function $C^{(0)}_g$ does not depend on $\mu_F$, whereas the low $x$
gluon distribution $F_g$ depends strongly on the scale. In other
words, based on \eqref{eq:LO}, the exclusive $V$ data measure
$g(x,\mu_F)$, but we do not know the value of the scale at which it
has been determined.  How does this scale freedom arise? To obtain the
LO result the coefficient function -- the upper box in
Fig.~\ref{fig:LO}(a) -- is calculated with on-mass-shell gluons with
transverse momenta $l_t=0$. In this collinear approach the
factorisation scale $\mu_F$ acts effectively as the ultraviolet (UV)
cutoff of the logarithmic integral $\int dl^2_t/l^2_t\propto
\ln\mu^2_F$ over the gluon transverse momentum in the gluon loop of
Fig.~\ref{fig:LO}(a). Formally, in the LO collinear factorisation
approach the value of $\mu_F$ is not known. In principle, the full result
does not depend on $\mu_F$ since the higher-order, NLO, NNLO, ...,
corrections compensate the effect of variations of $\mu_F$. However,
in reality the perturbative series is truncated and the compensation
may not be sufficient to provide the scale stability of the
theoretical prediction. 

On the other hand, we can go beyond the collinear logarithmic
approximation by computing the $l_t$ integral accounting for the $l_t$
dependence\footnote{A precise integration over $l_t$ was, in
  particular, implemented in~\cite{JMRT}.} of the hard $\gamma+gg\to
V$ matrix element, $\cal{M}$, shown in the upper box of
Fig.~\ref{fig:LO}(a). In comparison with the coefficient function
(calculated with $l_t=0$) now the $l_t$ and $l^2$ dependence of
$\cal{M}$ is included explicitly, and provides the UV convergence of
the integral over $l_t$. Formally, in collinear factorisation the
difference between the pure logarithmic ($\ln\mu_F^2$) evaluation and
the precise calculation of the gluon $l_t$ integral is treated as 
part of the NLO correction. This part of the NLO correction is of
kinematic origin and is usually quite large. Fortunately, it can be
moved into the LO component of the amplitude, noticeably reducing the
remaining NLO correction. 

Instead of performing an independent calculation which accounts for
the $l_t$ dependence, we can remain within the collinear approach
and determine the value of the $l_t$ integral given that the NLO
coefficient function $C^{(1)}_q$ of Fig.~\ref{fig:LO}(b) is
known. Indeed, Fig.~\ref{fig:LO}(b) is the only diagram for the quark
NLO coefficient function. In this approach the incoming quarks are
assumed to be on-mass-shell and with zero transverse momenta but the
loop integral over $l$, which contains the $l_t$ dependence, is
calculated exactly. Since this is the same integral as that which
occurs in Fig.~\ref{fig:LO}(a) we can use the result for $C^{(1)}_q$
to obtain a precise value, $J$, of the corresponding integral in the
LO amplitude of Fig.~\ref{fig:LO}(a). After this we choose a  scale
$\mu_F=\mu_0$ which mimics the precise $l_t$ integration. That is,
with a scale choice satisfying $\ln\mu^2_0=J$ we have moved a large
contribution from NLO to LO, and can continue to work in the
conventional collinear approach, but now with a smaller NLO
correction. 

\subsection{Transfer of part of the NLO to the LO contribution \label{sec:NLO}}
Since the above observation is crucial, let us demonstrate the
procedure in more detail. At NLO level the  LO+NLO amplitude at some
factorisation scale $\mu_f$ may be expressed in the form\footnote{For
  ease of understanding we omit the parton labels $a=g,q$ on the
  quantities in~(\ref{eq:sta}) and the following equations. The matrix
  form of the equations is implied.} 
\begin{equation}
A^{(0)}(\mu_f) + A^{(1)}(\mu_f) ~=~ C^{(0)} \otimes F(\mu_f) +
\alpha_s C^{(1)}(\mu_f) \otimes F(\mu_f)\,, 
\label{eq:sta}
\end{equation}
where the $F$s are the GPDs and where the coefficient function
$C^{(0)}$ does not depend on the factorisation scale. Note that we are
free to evaluate the LO contribution at a different scale $\mu_F$,
since the resulting effect can be {\it compensated} by changes in the
NLO coefficient function, which then also becomes dependent on
$\mu_F$. Then eq.~(\ref{eq:sta}) becomes 
\begin{equation}
A^{(0)}(\mu_f) + A^{(1)}(\mu_f) ~=~ C^{(0)} \otimes F(\mu_F) +
\alpha_s C^{(1)}_{\rm rem}(\mu_F) \otimes F(\mu_f)\,. 
\label{eq:stab1}
\end{equation}
Note that although the first and second terms on the right hand side
depend on $\mu_F$, their sum does not (to ${\cal O}(\alpha_s^2)$) and
is equal to the full LO+NLO amplitude calculated at the factorisation
scale $\mu_f$. 

In~(\ref{eq:sta}) the NLO coefficient function $C^{(1)}$ is calculated
from Feynman diagrams which are independent of the factorisation
scale. How does the $\mu_F$ dependence of $C^{(1)}_{\rm rem}$
in~(\ref{eq:stab1}) actually arise? It occurs because we must subtract 
from $C^{(1)}$ the $\alpha_s$ term which was already included in the
LO contribution.\footnote{Simultaneously this subtraction also provides
  the infrared convergence of $C^{(1)}$.} Since the LO contribution
was calculated up to some scale $\mu_F$ the value of $C^{(1)}$ after
the subtraction depends on the value $\mu_F$ chosen for the LO
component. The change of scale of the LO contribution from $\mu_f$ to
$\mu_F$ also means we have had to change the factorisation scale which
enters the coefficient function $C^{(1)}$ from $\mu_f$ to $\mu_F$. The
effect of this scale change is driven by the LO DGLAP evolution, which
is given by
\be
A^{(0)}(\mu_F)~=~
\left(C^{(0)} +\frac{\alpha_s}{2\pi}
    \ln\left(\frac{\mu_F^2}{\mu_f^2}\right) C^{(0)} \otimes V \right)
\otimes F(\mu_f)\,,
\label{eq:dg}
\ee
where $V$ denotes the skewed splitting functions. That is, by choosing
to evaluate $A^{(0)}$ at scale $\mu_F$ we have moved the part of the NLO
(i.e. $\alpha_s$) corrections given by the last term of~(\ref{eq:dg})
from the NLO to the LO part of the amplitude. In this way $C^{(1)}$
becomes the remaining $\mu_F$-dependent coefficient function
$C^{(1)}_{\rm rem}(\mu_F)$ of~(\ref{eq:stab1}). In spite of the unusual form of (\ref{eq:stab1}), with two different scales $\mu_f$ and $\mu_F$, it is an exact equality at NLO and could, in principle, be generalised to higher orders.\footnote{For example, if the NNLO contribution were known, then we will have three scales: $\mu_f$, $\mu_F\equiv \mu^{\rm LO}$, and $\mu^{\rm NLO}$, where the NNLO correction to  (\ref{eq:stab1}) takes the form $\alpha_s^2 C^{(2)}_{\rm rem}(\mu^{\rm LO},\mu^{\rm NLO})\otimes F(\mu_f)$. The scale $\mu^{\rm NLO}$ is fixed to nullify this term in the limit $x \gg \xi$, and hence further reduce the sensitivity to variations of the scale $\mu_f$.}

\subsection{Large $\ln(1/\xi)$ terms resummed in the LO contribution}
The idea is to use the above procedure to reduce the scale dependence
of the exclusive $V$ photoproduction amplitude. Unfortunately the
value of $\mu_F$ is just one number, while $C^{(1)}(x/\xi)$ is a
function of the ratio $x/\xi$. So we have no chance to nullify
$C^{(1)}$ completely. Nevertheless we can nullify the most important
NLO correction which is enhanced by the large value of
$\ln(1/\xi)$. This contribution is generated by the integral $\int
(dx/x)$ in the $C^{(1)}_{\rm rem}(x/\xi,\mu_F)  F(x,\xi,\mu_f)$
convolution of eq.~(\ref{eq:stab1}) over the kinematic region of $1\gg
x\gg \xi$. That is, we choose a scale $\mu_F=\mu_0$ which nullifies
$C^{(1)}(x/\xi)$ in the limit of $x\gg \xi$. It can be seen
from~(\ref{eq:he}) that the scale $\mu_F=\mu_0=m$ ensures that the
$\ln(1/\xi)$-enhanced NLO corrections completely vanish for both the
quark and the gluon components. 

From the NLO viewpoint the particular choice of $\mu_F=m$ in the LO
part is irrelevant. The corresponding order $\alpha_s$ effect is
exactly compensated by the remaining $C^{(1)}_{\rm rem}$
term. However, the variation of $\mu_F$ in the LO part affects not
only the ${\cal O}(\alpha_s)$ terms but the higher-order  $\alpha_s$
contributions as well. Therefore, in this way, we resum the important
large $\ln(1/\xi)$-enhanced part of the higher-order $\alpha_s$
corrections inside the parton distribution convoluted with the LO
coefficient function and improve the convergence of the perturbative
series.\footnote{In particular, the most important factorisation scale
  dependence, enhanced by large $\ln(1/\xi)$, is caused by the double
  log terms, $[\alpha_s\ln(\mu_F)\ln(1/x)]^n$, generated in the axial gauge by ladder-type diagrams. For GPDs this ladder contribution was studied in \cite{BL}, where it was shown,           
   in the large $\ln(1/x)$ limit, that the
  skewed splitting functions are proportional to $1/x$.  So these double log terms come from DGLAP
  integrals of the form 
$$
\left(\int^{\mu_F^2}\frac{dk^2_n}{k^2_n}
\int^{k_n^2}\frac{dk^2_{n-1}}{k^2_{n-1}} \ldots
\int^{k_2^2}\frac{dk^2_1}{k^2_1}\right)~\left(\int^1_x\frac{dx_n}{x_n} 
  \int^1_{x_n}\frac{dx_{n-1}}{x_{n-1}}
  \ldots \int^1_{x_2}\frac{dx_1}{x_1}\right) ~\sim ~
[\alpha_s\ln(\mu_F)\ln(1/x)]^n/(n!)^2
$$ 
where we have strong $k^2 \equiv k^2_t$ and $x$ ordering. Thanks to
the strong $k_t$ and $x$ ordering in these double log LO integrals,
the correct upper limit, $\mu_F$, in the first integral automatically
provides the exact resummation of all the terms in the double log
series.} 

Actually our approach is rather close in spirit to the
$k_t$-factorisation method. Indeed, there, the value of the 
factorisation scale is driven by the structure of the $k_t=l_t$ (or the
virtuality, $Q^2$) integral in the diagrams of
Fig.~\ref{fig:LO}. 
\footnote{We stress again that, in the high energy ($x\gg \xi$) contribution, the form of the integral over the gluon loop momentum $l_t$ is exactly the same in both the quark (Fig.~\ref{fig:LO}(b)) and gluon (Fig.~\ref{fig:LO}(b) but with the quark lines replaced by gluons) NLO contributions. Therefore the scale $\mu_F=\mu_0$ simultaneously nullifies the high energy quark and gluon contributions. Note that this is only true after the corrections discussed in Section~\ref{sec:NLOgen} are included.}
In the $k_t$-factorisation approach this $k_t$
integral is written explicitly, while the parton distribution {\it
  unintegrated} over $k_t$ is generated by the last step of the DGLAP
evolution, similar to the prescription proposed in
Refs.~\cite{KimbMR,WMR}. Now, using the known NLO result, we account
for the {\em exact} $k_t$ integration in the last cell adjacent to the
LO hard matrix element. This hard matrix element $\cal{M}$, shown in
the upper box of Fig.~\ref{fig:LO}(a), provides the convergence of the
integral at large $k_t$. In this way it puts an effective upper limit
of the $k_t$ integral, which plays the role of an appropriate
factorisation scale. 

The details of the prescription, for the case of the high-energy
Drell-Yan process, were discussed in~\cite{DY}. Indeed, Drell-Yan
production of low-mass lepton pairs at high rapidity is another
process for which the NLO prediction depends sensitively on the choice
of the factorisation scale, unless the $\ln(1/x)$ enhancements are
first resummed in the incoming parton distributions. 
It was found in Ref.~\cite{DY} that, after the scale $\mu_F=\mu_0$ is
fixed for the LO contribution, the variation of the scale in the
remaining NLO part does not noticeably change the predicted Drell-Yan
cross section. In~\cite{DY} it was shown how to calculate $\mu_0$ for
the Drell-Yan process and, moreover, that the NLO prediction with
$\mu_F=\mu_0$ is very close to the NNLO result. 

Returning to the earlier discussion in this subsection, it is clear
from~(\ref{eq:he}) and~(\ref{eq:stab1}) that the scale choice
$\mu_F=m=M_V/2$ provides the appropriate resummation of the
$\ln(1/\xi)$ enhanced terms, and, as a consequence, suppresses the
remaining high-energy NLO corrections. In other words, the GPDs with
$\mu_F=m$  (chosen in the LO part) include all the $\ln(1/\xi)$
enhanced contributions, while the remaining NLO corrections arise only
from hard matrix elements corresponding to intermediate states of
relatively small mass of about $M_V$. 

\begin{figure} [t]
\begin{center}
\includegraphics[width=0.45\textwidth]{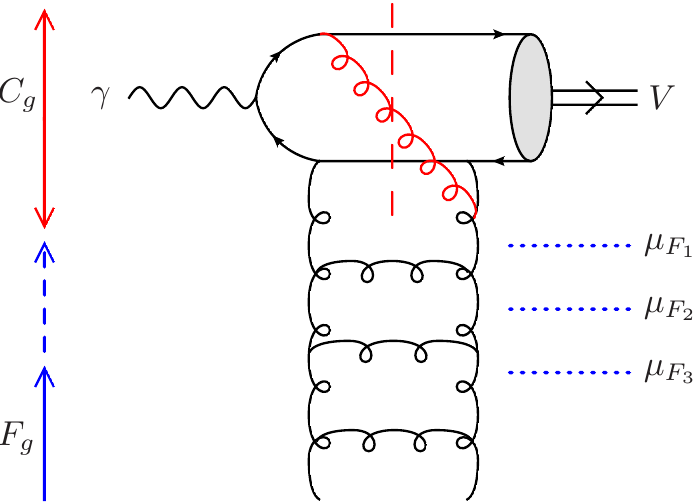}
\caption{\sf A Feynman diagram which illustrates different
  possibilities for the division of cells between the parton evolution
  and the hard matrix element, dependent on the choice of the
  factorisation scale, $\mu_{Fi}$.} 
\label{fig:2}
\end{center}
\end{figure}

We can also explain this pictorially.  As noted above, the $\mu_F$
dependence of the NLO correction is caused by the subtraction of the
contribution generated by the evolution equation. In particular, the
correction which (i) depends on $\mu_F$ and (ii) is enhanced by
$\ln(1/\xi)$, is  generated by the ladder-type
diagrams\footnote{Ladder diagrams occur in the axial gauge which is
  conventionally used to calculate the GPDs.} shown in
Fig.~\ref{fig:2}. The choice of the factorisation scale $\mu_{Fi}$
determines which part of the  diagram is attributed to the evolution
of the PDF and which to the hard matrix element. By choosing the value
of $\mu_F=\mu_0=M_V/2$ we include all the ladder cells in the LO part,
so that the remaining NLO matrix element will not receive any
contribution from ladder-type diagrams (at small $\xi$). Note that the
ladder diagrams would contribute to the NLO matrix element (from
$\mu_F$ to $\mu_0$) if we were to choose a scale $\mu_F < \mu_0$. For
the choice $\mu_{F} > \mu_0$ the NLO matrix element would contain the
ladder diagrams with the negative sign in order to compensate the
extra contributions (from $\mu_0$ to $\mu_F$) included in the LO part. 

Since we consider an exclusive process, the mass of the final state is
fixed, but the intermediate states, corresponding to the discontinuity
which gives the imaginary part of the amplitude (depicted using a dotted
line in Fig.~\ref{fig:2}), depend on the choice of scale. For larger
values of $\mu_F$ the mass of the allowed intermediate states in the
matrix element becomes smaller. Of course, it is impossible to absorb
inside the evolution all intermediate states other than the heavy quark pair
$Q{\bar Q}$ state which produces the $V$; for example we cannot absorb
the effects of the uppermost gluon in Fig.~\ref{fig:2} emitted by the
upper heavy quark. Nevertheless, when we choose the appropriate
scale $\mu_F=M_{V}/2$, the mass of the remaining intermediate states
becomes of the order of $M_V$. This choice will move $\ln(1/\xi)$
contributions from the matrix element into the GPD. It will therefore
provide a much more stable final result, since the remaining NLO
contributions then cannot be enhanced by the large values of
$\ln(1/\xi)$. 

It should be emphasised that the asymptotics of the NLO amplitude is
used only to determine the effective scale $\mu_F$. In all our further
numerics we use the full expression for the complete NLO amplitude as
given in~\cite{Ivan,SJonesThesis, GJT}.

\section{Can exclusive vector meson production data be included in a
  global PDF fit?} 
In this section we present our results for the exclusive $\gamma p
\to J/\psi +p$ and $\gamma p \to \Upsilon +p$ processes as functions
of the factorisation scale $\mu_f$, and the renormalisation scale
$\mu_R$. The cross section relies sensitively on the gluon PDF at
small $x$, in a kinematic regime where it is very poorly known, as
well as depending of the choice of scales. For these quasi-elastic
processes we present just the dominant imaginary part of the
amplitude. The real part of the coefficient functions has been calculated exactly, both via
a dispersion relation \cite{Ivan} and by directly computing the real part of the loop integrals \cite{SJonesThesis,GJT}, it is non-zero in the time-like
region $|x|<\xi$. Therefore, to compute the real part of the amplitude directly, we need also
the GPDs in this region. Unfortunately, the Shuvaev transform is not valid in the time-like 
region \cite{ShuvNockles}. Nevertheless, the real part of the amplitude can
be included via dispersive methods on the level of the amplitude. However, giving a
prediction of the full cross section is not our objective here. 
Rather our aim is to study the stability of the perturbative
predictions and to investigate whether or not we can determine optimum
scales so that experimental data for these processes, and the related
$pp\to p\,V\,p$ processes, can be included in global PDF
(collinear) analyses to constrain the gluon PDF at low $x$, in a
domain for which, at present, there are no data. 
For this goal it is sufficient to work with the more simple imaginary part of the amplitude.

We recall the main result of Section 2. We start from the key
equation,~(\ref{eq:stab1}), that is 
\begin{equation}
A^{(0)}(\mu_f) + A^{(1)}(\mu_f)~=~ C^{(0)} \otimes F(\mu_F) + \alpha_s
C^{(1)}_{\rm rem}(\mu_F) \otimes F(\mu_f)\,. 
\label{eq:stab11}
\end{equation}
To obtain this result we had introduced by hand a new scale $\mu_F$ in
the LO term and showed that the choice $\mu_F=m= M_V/2$ allowed us to
resum and transfer all the enhanced large $\ln(1/\xi)$ terms from the
NLO contribution to the LO term, leaving a much smaller remaining NLO
contribution $\alpha_s C^{(1)}_{\rm rem}(\mu_F) \otimes
F(\mu_f)$. There is still a $\mu_f$ scale dependence, but now this should
be much weaker.

\subsection{The process $\gamma p \to J/\psi +p$  \label{sec:J}}
\begin{figure} 
\begin{center}
\includegraphics[width=0.35\textwidth,angle=-90]{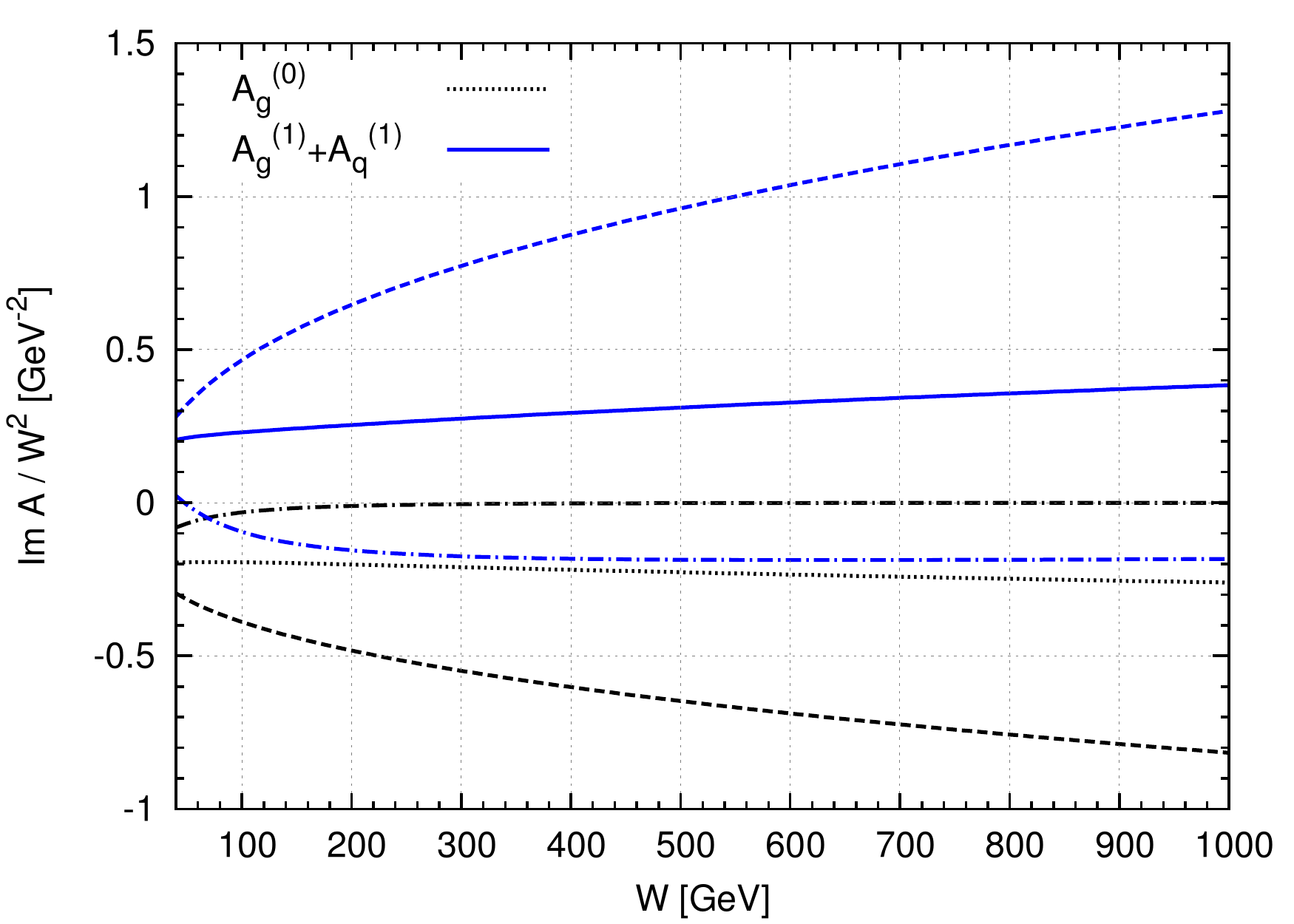}
\includegraphics[width=0.35\textwidth,angle=-90]{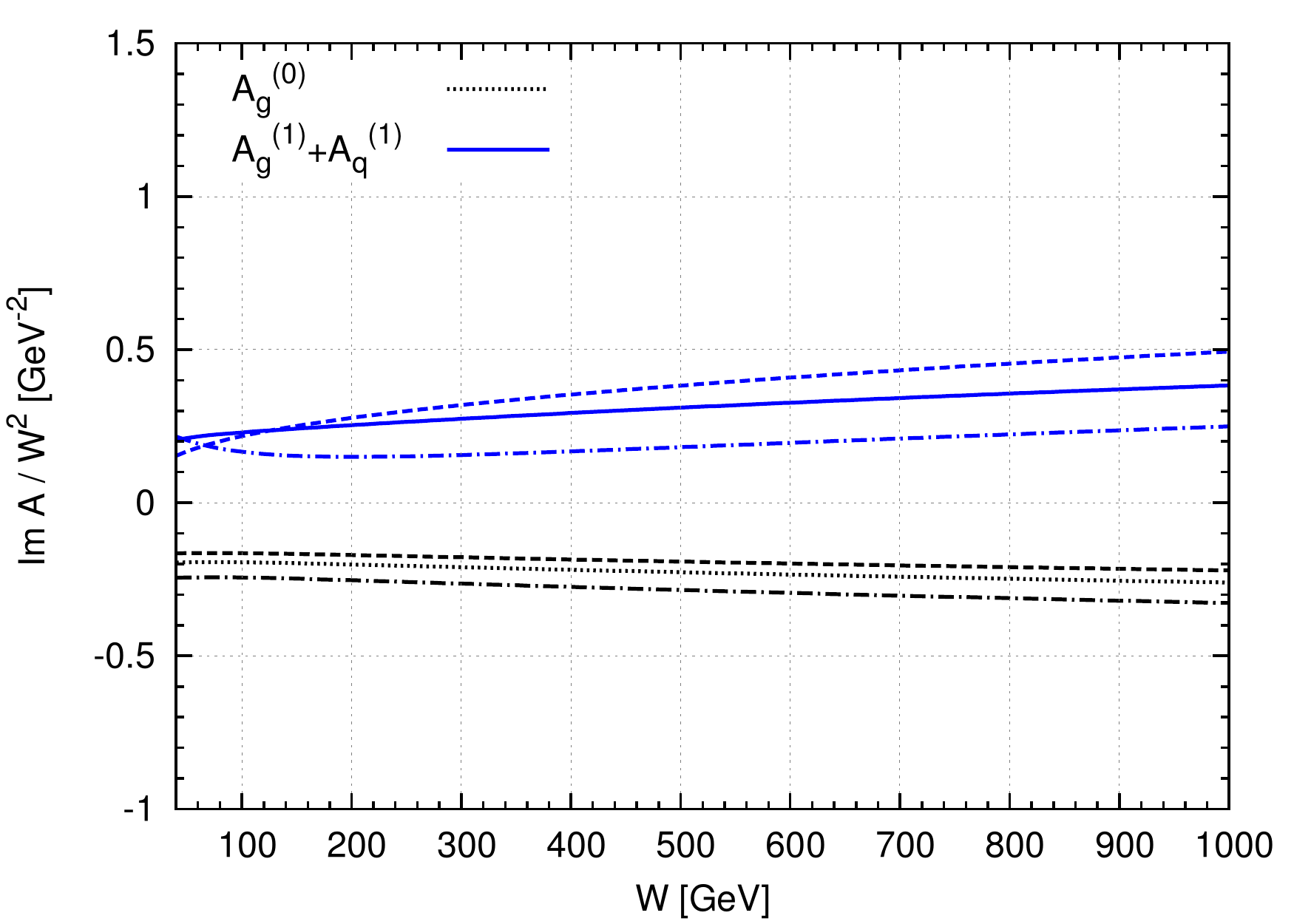}
\caption{\sf 
Predictions of $\mathrm{Im} A/W^2$ for $\gamma p \rightarrow \J + p$
as a function of the $\gamma p$ centre-of-mass energy $W$, 
produced using CTEQ6.6 partons~\cite{Nadolsky:2008zw} with scales
$\mu \equiv \mu_F = \mu_f = \mu_R$ (left panel) and $\mu \equiv \mu_f
= \mu_R$ with $\mu_F^2 = 2.4~\GeV^2$ (right panel). The bottom (top)
set of curves corresponds to the Born (1-loop) contribution with the
scale variation $\mu^2 = 1.2, 2.4, 4.8~\GeV^2$. The dot-dashed, solid
and dashed lines correspond to the low, central and high values of the
scale $\mu$, respectively. Note that for the left panel, the bands
overlap for energies bigger than about $70~\GeV$.} 
\label{fig:J}
\end{center}
\end{figure}

To obtain predictions for exclusive $J/\psi$ photoproduction we are
working at scales close to the input GPDs used in the calculation. The
results for the scale variation of the imaginary part of the amplitude
are shown in the two plots of Fig.~\ref{fig:J}.  In each plot we show
separately the LO and NLO contributions to the amplitude.\footnote{We
  choose to use CTEQ66 partons~\cite{Nadolsky:2008zw} since the gluon
  obtained is positive definite and since they were used in our
  earlier works on the subject~\cite{MNRT,JMRT}. Moreover, since we
  are emphasising the general procedure rather than making
  quantitative predictions, the choice of any particular, reasonable
  PDF set is not important.} The left plot shows how these
contributions change if we vary all the scales
$\mu_f=\mu_F=\mu_R\equiv \mu$ simultaneously, taking
$\mu^2=m^2/2,~m^2,~2m^2$.  In the right plot we fix $\mu_F$ at the
optimum scale $\mu_F=m$ and vary $\mu_f=\mu_R\equiv \mu$. The result
is dramatic.  The transfer of the $\ln(1/\xi)$ terms from the NLO to
the LO contribution has significantly reduced the $\mu_f$ scale
dependence. 

\begin{figure}
\begin{center}
\includegraphics[width=0.6\textwidth]{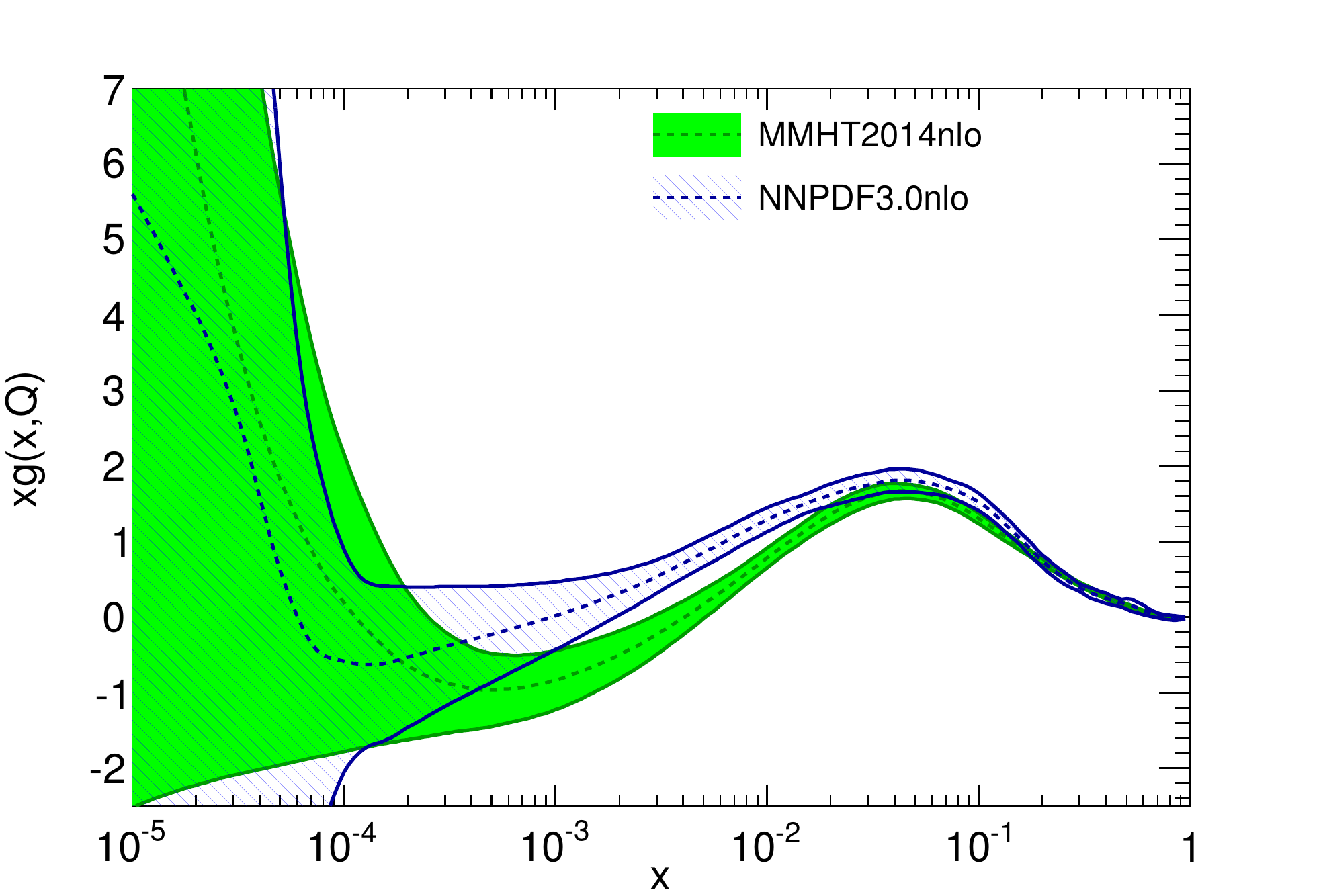}
\caption{\sf The gluon distribution at $Q^2=1.21~\GeV^2$ as determined
  by two recent global
  analyses~\cite{Ball:2014uwa,Harland-Lang:2014zoa}. Figure produced using~\cite{Bertone:2013vaa,Carrazza:2014gfa}.}  
\label{fig:lowxgluon}
\end{center}
\end{figure}

But there is another, more severe problem. The LO contribution is
dominated by the NLO contribution of opposite sign. Thus the imaginary
part of the quasi-elastic $\gamma p \to J/\psi +p$ amplitude changes
sign when the NLO contribution is added. As it stands, this result is
in contradiction with modelling the interaction as an elastic forward
scattering where the imaginary part of the amplitude is positive. What
is happening? The explanation is interesting. In general, the global
DGLAP PDF analyses start from input forms which are completely
arbitrary and, moreover, neglecting any constraints, know nothing
about the structure of the evolution at low $Q^2$. It is then found
that the gluon PDF tends to be small, or even negative, in the low
$Q^2$, low $x$ $(10^{-4}\lapproxeq x \lapproxeq 10^{-2})$ domain, see
Fig.~\ref{fig:lowxgluon}. Clearly due to the lack of data constraints
in this domain the gluon is not reliably known. 

Let us study the over-simplified case with an input gluon $g(x)=0$ so
that at the input scale we have only quark PDFs. For $\gamma p \to
J/\psi +p$ the quark contribution only appears at NLO. The imaginary
part of this NLO contribution is negative with respect to the normal
$(g \ne 0)$ LO term. Indeed, when calculating the NLO coefficient
function we must subtract the contribution which is already generated
by LO evolution. However, the subtraction is performed purely formally
in order to avoid the infrared divergence and {\em does not} account 
for the actual value of the gluon density used to calculate the LO term.
Therefore, at low scales, where the LO term is suppressed by much too
small `unphysical' gluon PDFs obtained from the global fits, the NLO
term generated by the quark is not just a correction, but is the
dominant contribution. As a consequence, the imaginary part of the
quasi-elastic amplitude becomes negative (in comparison with the LO
contribution)  and the observed behaviour reveals a lack of stability
of the perturbative series at this order. Of course, at high scales
where evolution, and not the input, determine the PDFs, we obtain a
sensible NLO prediction. 

This simple consideration demonstrates that if $\gamma p \to J/\psi
+p$ and $pp\to p + J/\psi + p$ data could be included in a global fit,
then they would put a strong constraint on the low $x$ input gluon
distribution. It is not simply that $g(x)$ must be positive, but that
actually the input gluons cannot be smaller than the density of gluons
emitted by the quarks before the beginning of the evolution, when the
parton virtuality $Q<Q_0$. If $J/\psi$ data would have been
included in the collinear global parton analyses, this constraint on
the low $x$ input gluon would have been automatically
satisfied. 

Indeed, it is seen from the second plot of Fig.~\ref{fig:J} that after
the $\ln(1/\xi)$ enhanced corrections are resummed by fixing
$\mu_F=m$, the stability of both the LO and the NLO components of the
amplitude under the scale variations are much better. From this
viewpoint the $J/\psi$ data may be included in a global PDF analysis.
The only problem is that the gluon density obtained in the existing
PDF analyses is too small at input scales and low $x$
($10^{-5}\lapproxeq x \lapproxeq 10^{-2}$), so that we get the wrong
sign of the imaginary part of the amplitude.  This means that if
$J/\psi$ data were included in the global PDF analyses, we must get a
larger gluon density at low scales.\footnote{Most probably the pure
  DGLAP global analyses should take into account  absorptive
  corrections, which are not negligible at low $x$ and low $Q^2$. The
  existing very small gluon densities in this kinematic domain are a
  way of mimicking these absorptive effects.} Larger gluons in this
kinematic domain will increase the LO component of the $J/\psi$
photoproduction amplitude and will provide the correct sign for the
whole LO+NLO amplitude. 

\subsection{The processes $\gamma p \to \Upsilon +p$ and $pp\to
  p\Upsilon p$  \label{sec:upsilon}} 
\begin{figure}
\begin{center}
\includegraphics[width=0.35\textwidth,angle=-90]{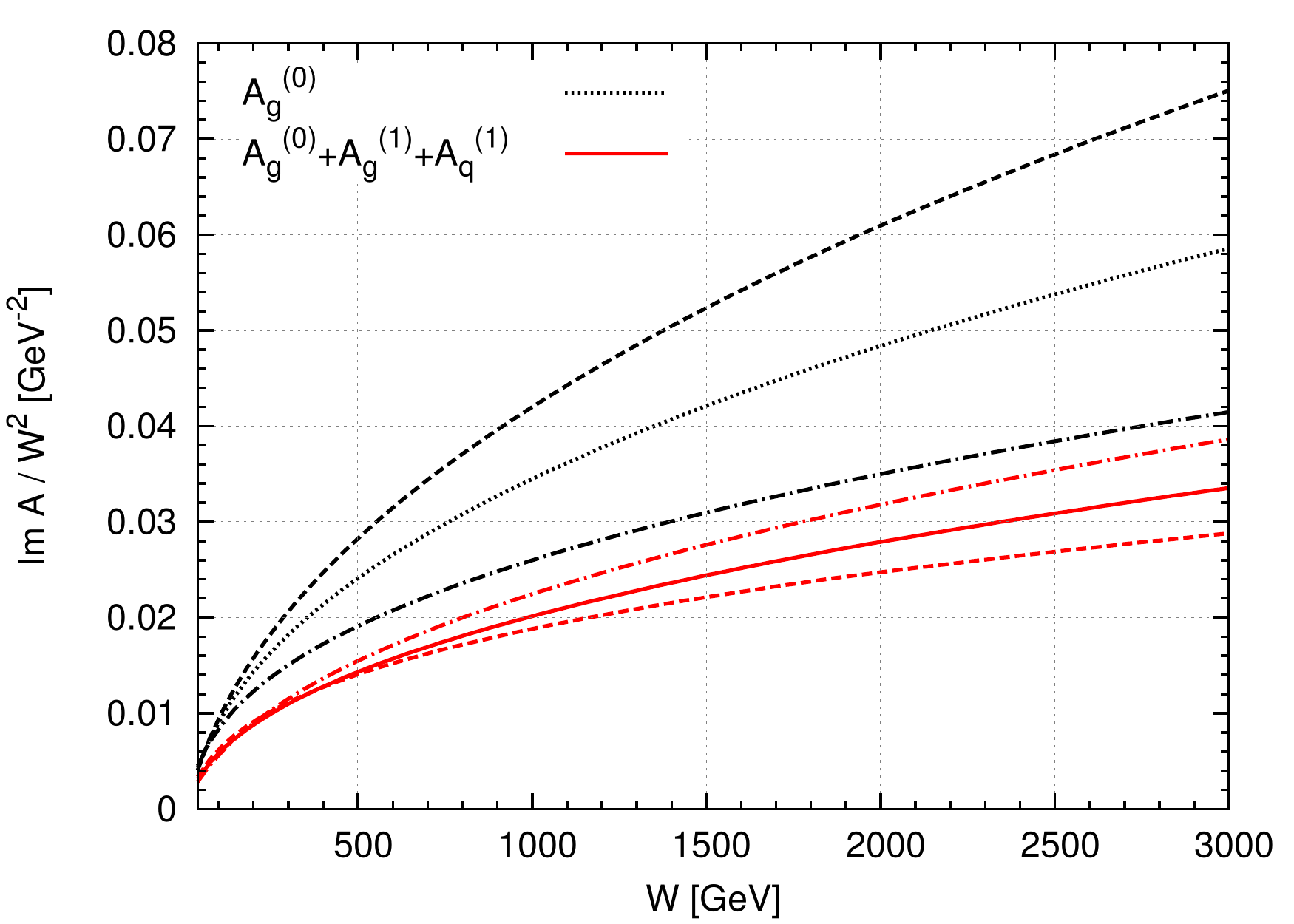}
\includegraphics[width=0.35\textwidth,angle=-90]{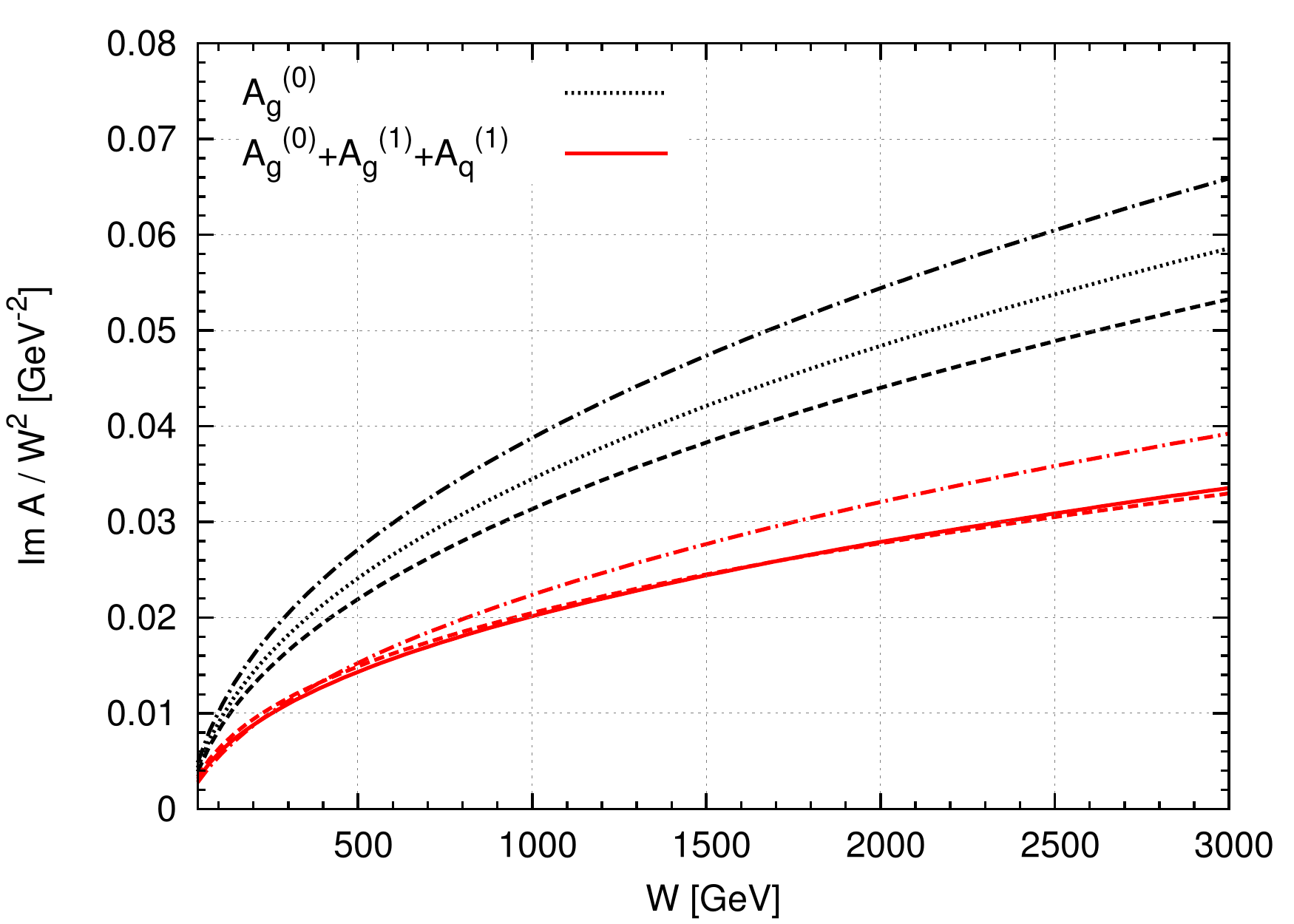}
\includegraphics[width=0.35\textwidth,angle=-90]{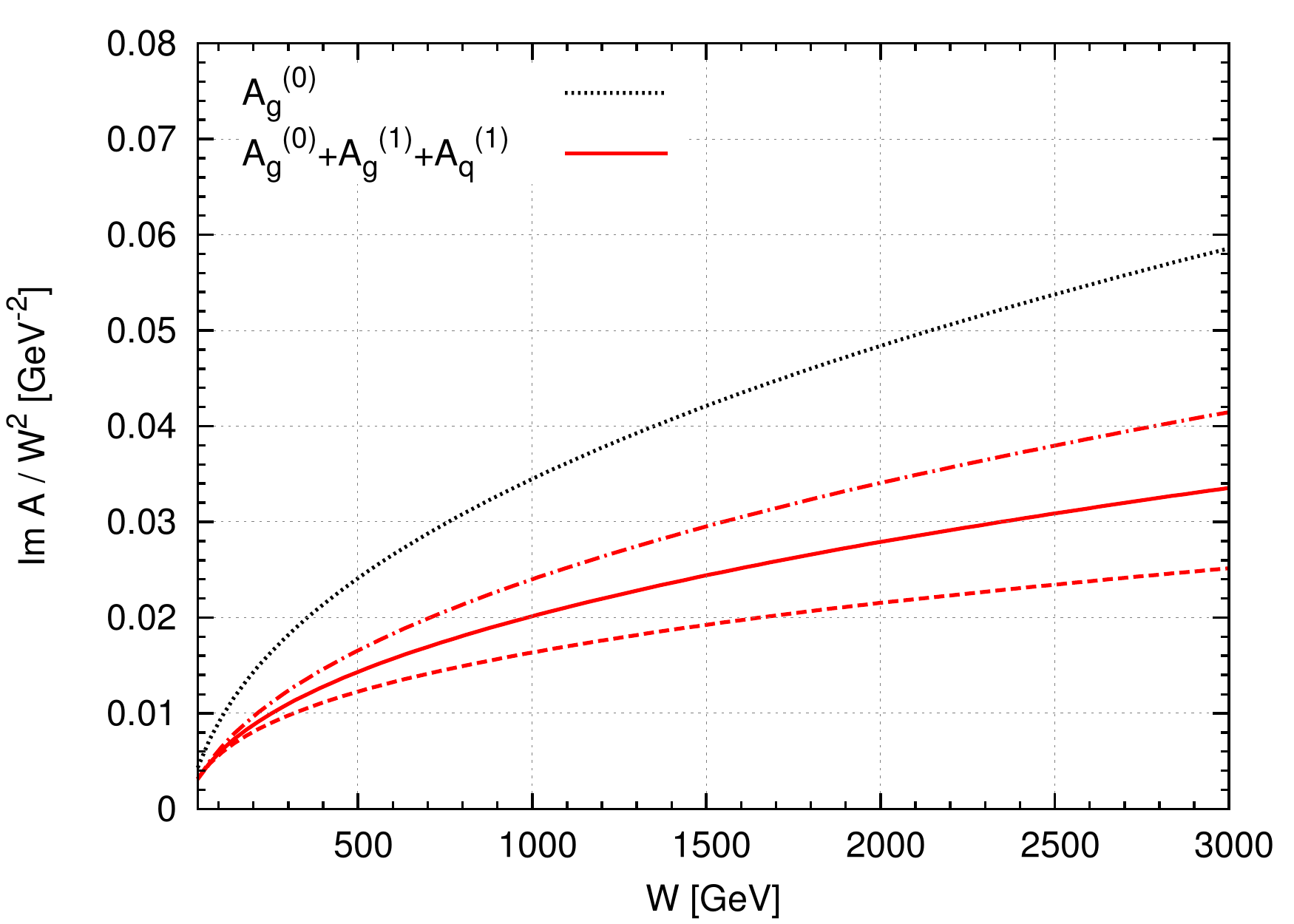} 
\includegraphics[width=0.35\textwidth,angle=-90]{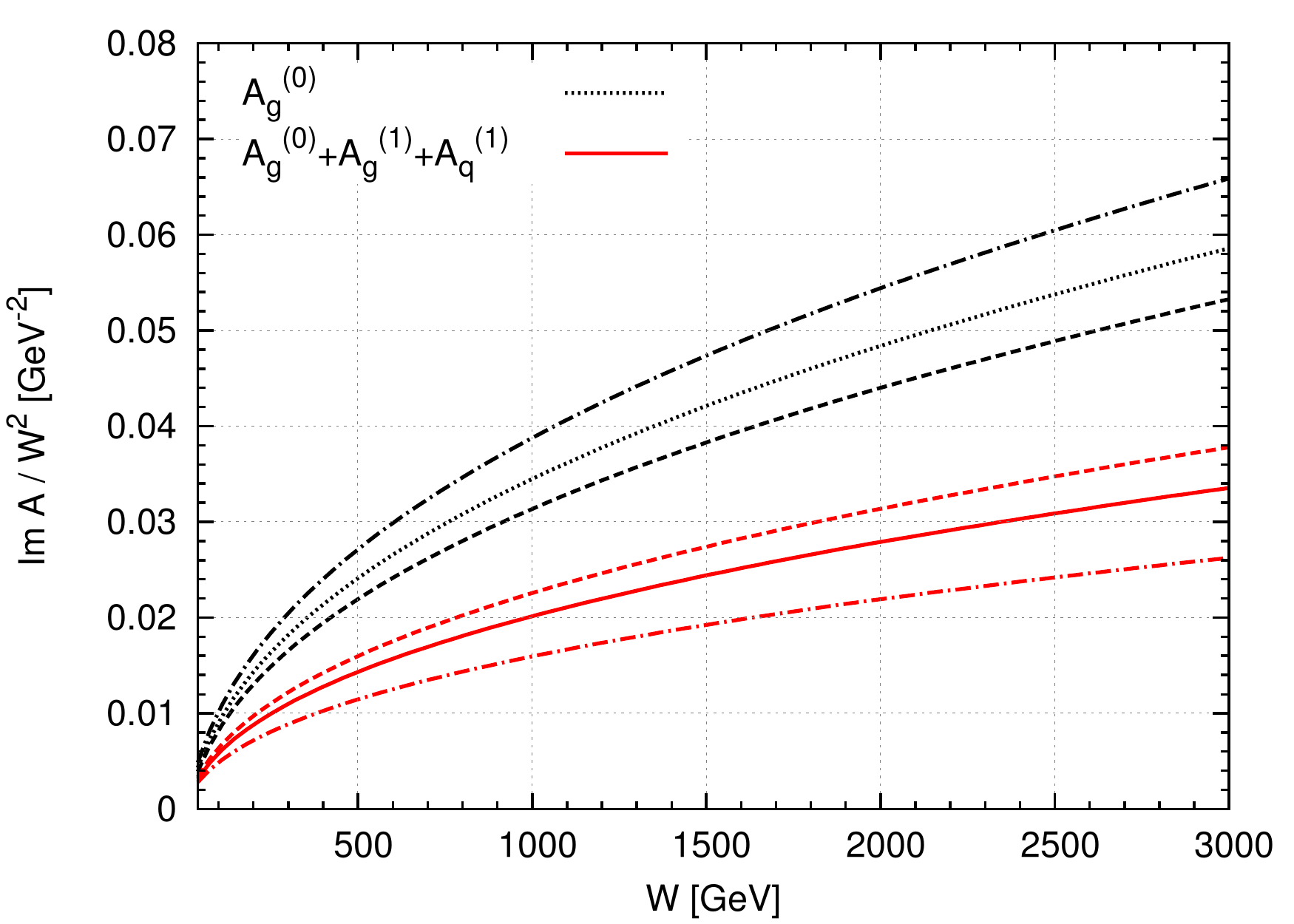} 
\caption{\sf  
Predictions of $\mathrm{Im} A/W^2$ for $\gamma p \rightarrow \Upsilon
+ p$ as a function of the $\gamma p$ centre-of-mass energy $W$, 
produced using CTEQ6.6 partons~\cite{Nadolsky:2008zw}. The scales
are $\mu \equiv \mu_F = \mu_f = \mu_R$ (top left), $\mu \equiv \mu_f =
\mu_R$ with $\mu_F^2 = 22.4~\GeV^2$ (top right), $\mu \equiv \mu_f$
with $\mu_F^2 = \mu_R^2 = 22.4~\GeV^2$ (bottom left), $\mu \equiv
\mu_R$ with $\mu_f^2 = \mu_F^2  = 22.4~\GeV^2$ (bottom right). The top
(bottom) set of curves corresponds to the LO (LO+NLO) contribution
with the scale variation $\mu^2 = 11.9, 22.4, 44.7~\GeV^2$. The
dot-dashed, solid and dashed lines correspond to the low, central and
high values of the scale $\mu$, respectively. The large GPD
uncertainties are not shown.} 
\label{fig:U1}
\end{center}
\end{figure}
\begin{figure}
\begin{center}
\includegraphics[width=0.4\textwidth]{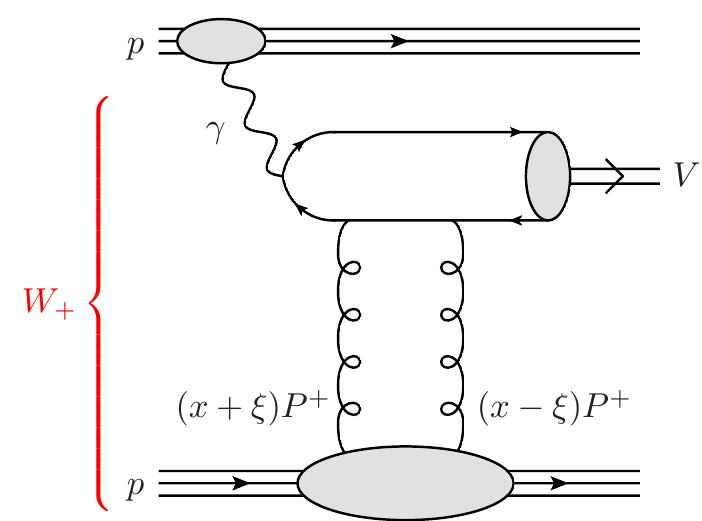}
\qquad
\includegraphics[width=0.4\textwidth]{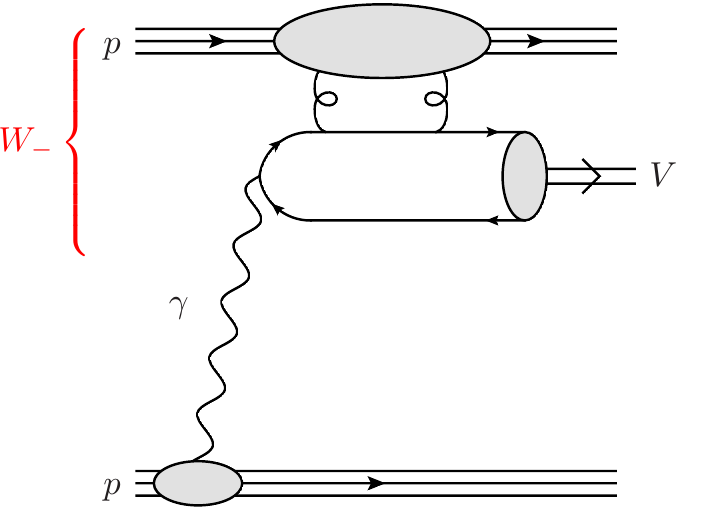}
\caption{\sf The two diagrams describing exclusive $\Upsilon$
  production at the LHC.  Diagram (a), the $W_+$ component, is the dominant
  contribution for an $\Upsilon$ produced at large rapidity $y$. Thus
  data for $pp\to p\Upsilon p$ allow a probe of very low $x$ values,
  $x\simeq M_\Upsilon ~e^{-y}/\sqrt{s}$; recall that
  in the (dominant) imaginary part of the Born amplitude we have $x=\xi$.} 
\label{fig:Udiag}
\end{center}
\end{figure}

The top two plots of Fig.~\ref{fig:U1} show the results for the scale
sensitivity of $\gamma p \to \Upsilon +p$, analogous to those of
Fig.~\ref{fig:J} for $\gamma p\to J/\psi +p$. Again we see that fixing
the scale $\mu_F=M_\Upsilon/2$ reduces the $\mu_f$ scale
uncertainty. The lower two plots show that part of this stability
arises because the change caused by  variation of $\mu_R$ is to some
extent compensated by an `opposite' change due to the variation of $\mu_f$.
If we were to choose a non-optimal scale, say for example,
$\mu^2_F=2m^2$, then the scale variation turns out to be about twice
as large as for the optimal choice $\mu^2_F=m^2$.
Due to the larger $\Upsilon$ mass we are now working at scales with
more perturbative stability. The LO contribution is partly cancelled
by the NLO term, but is not dominated by it. If we take the upper
right or lower left plot of Fig.~\ref{fig:U1} then the $\mu_f$ scale
uncertainty of the amplitude is about $\pm 15\%$ and $\pm 25\%$
respectively. As a result the data for exclusive $\Upsilon$ production
can be used in global PDF analyses to probe the gluon distribution
down to $x \sim 10^{-5}$, in the case of LHCb kinematics. The
calculation of exclusive $pp\to p\Upsilon p$ is described
in~\cite{JMRT}. It is based on the sum of the two diagrams shown in
Fig.~\ref{fig:Udiag}. For an $\Upsilon$ produced at large rapidity
$y$, the dominant contribution is from the diagram with the larger
$\gamma p$ centre-of-mass energy $W_+$, which depends on the gluon
density at $x\simeq M_\Upsilon ~ e^{-y}/\sqrt{s}$. The small
contribution from the other diagram, with much lower energy $W_-$, may
be estimated from the existing HERA data. 

If we use the central values of the presently available PDF sets
obtained from global analyses, then we find cross section
predictions which are about a factor of four below the HERA
exclusive $\Upsilon$ data~\cite{H1:3,ZEUS:6,ZEUS:7}. This indicates
that  future PDF global analyses with $\Upsilon$ data included  will,
just as for $J/\psi$ data, require a larger gluon distribution at low
values of $x$. While the exclusive $\Upsilon$ process samples the PDFs
at scales $Q^2\sim M_\Upsilon^2 /4$, the larger gluon required at
these $Q^2$ will affect the gluon densities at all values of $Q^2$ via
the evolution.

\subsection{Note on the alternative $k_t$-factorisation approach}
The collinear DGLAP global analyses tend to result in gluon
distributions with valence-like $x$ distributions at low scales, see
Fig.~\ref{fig:lowxgluon}. On the other hand, the approach used in the
JMRT paper~\cite{JMRT}, to study  exclusive vector meson production,
is free from this problem. There, the `NLO' prediction was not
calculated as a correction from NLO Feynman diagrams in collinear
factorisation, but approximated by taking the full integral over the
gluon $k_t$ in Fig.~\ref{fig:LO}(a), including an ansatz for the now
$k_t$ dependent gluon. The convergence of this explicit $k_t$ integral
provides effectively the `optimal' value of the factorisation scale
and in this way sums all $\ln(1/\xi)$ enhanced terms, accounting for a
large part of the NLO corrections. It does, of course, not include
corrections from diagrams which do not have a ladder structure,
however, the contribution from the light quarks, Fig.~\ref{fig:LO}(b), 
is already included in the unintegrated gluon distribution used in
JMRT~\cite{JMRT}. Clearly, then the quasi-elastic amplitude is positive
definite. Let us briefly describe how our `NLO' predictions for
$\gamma p \to \Upsilon +p$ data were made. First the incoming gluon
distribution was fitted, within this approach and using an ansatz for
the gluon based on the important double-logarithmic dependence on
$\ln(1/x)$ and $\ln Q^2$, to reproduce the $J/\psi$ data from HERA and
the LHC. Then we proceeded to $\Upsilon$ production using our fitted
gluon. We verified that, in the relevant kinematic domain, this gluon
reproduces the NLO DGLAP evolution to good accuracy. Therefore it was
not surprising that the `NLO' {\it predictions} for the $\Upsilon$
cross section, as a function of $W$, agreed well with the LHCb
data~\cite{LHCb:3} when they became available.

\section{Conclusions \label{sec:conclusions}}
Here we have been concerned about the inclusion of exclusive  vector meson production data in
 global PDF analyses in order to probe
 the gluon density at small values of $x$; that is, in
the $x\lapproxeq 10^{-4}$ domain. Note that in this domain the input gluon
PDF is freely parametrized, and is found to have a tendency to be
valence-like with large uncertainties. It was hoped that the situation
would be changed as data for exclusive vector meson production,
$V=J/\psi$ and $ \Upsilon$, which are very sensitive to the gluon at
small $x$, became available at HERA and the LHC; particularly
measurements of the exclusive process $pp\to p\,V\,p$ at the LHC with
$V$ detected at large rapidity. Why has this not happened (so far)? 

For $J/\psi$, the main problem is, at first sight, the very poor
convergence of the perturbative expansion in the collinear
approach. This can be seen, e.g., from Fig.~\ref{fig:J}, where the NLO
contribution is comparable to the LO term and opposite in sign. 
However, we argued that this large NLO contribution, in comparison
with that of the LO, reflects not the poor convergence, but rather
that the global PDF analyses find a gluon density which is too small
at low $x$ and low scales. For this reason we find a LO contribution
to  exclusive $J/\psi$ photoproduction amplitude which is too small.
We argued that  the input gluon in the collinear approach, used in the
global PDF analyses, should not be parametrized freely, but should be
subject to some constraints. We noted that working in the physical,
$k_t$-factorisation type scheme would avoid these problems. 

For exclusive $\Upsilon$ production the situation is much better. We
found that the optimum factorisation scale is much higher, the
perturbative expansion at NLO in collinear factorisation converges
well, and the remaining mild scale dependence of the predictions
($\sim \pm 15\%$ on amplitude level) means that data for $pp\to
p\Upsilon p$ can now be included in the global PDF fits to determine
the gluon in the low $x$ regime for the first time.

It is appropriate to list the theoretical uncertainties of the present calculation. Although the leading double log terms have been resummed correctly to all orders, there still exist remaining NNLO and higher contributions, which are unknown at present. When the full NNLO amplitude is known we showed how the scale uncertainty can be further reduced. Next, we consider the accuracy of the expressions which relate GPDs to conventional PDFs,  (\ref{eq:shuvq}) and (\ref{eq:shuvg}). These relations are based on conformal invariance and the equality of the Gegenbauer moments of GPDs to the Mellin moments of PDFs. Due to the polynominal property \cite{Ji} the accuracy of this equality of the moments is ${\cal O}(\xi^2)$, providing an ${\cal O}(\xi)$ accuracy for (\ref{eq:shuvq}) and (\ref{eq:shuvg}), see \cite{ShuvNockles}. 
However, recall that the Shuvaev transform {\em assumes} the absence of the singularities in the right-half
Mellin-$N$ plane  for the input distribution. This assumption is
physically reasonable since in the Regge approach there are no
singularities in the right-half ($j>1$) plane in the space-like
($|x|>\xi$) domain where we actually work.  Nevertheless, whenever possible, this assumption should be checked.  One check is that the predictions for the GPD/PDF ratio
obtained in this way are in a good agreement (especially for low $x$
gluons, which are the most important for exclusive $J/\psi$ or $\Upsilon$
production) with the NLO results of \cite{KM}
coming from a fit to deeply virtual Compton scattering (DVCS) HERA data. 
Finally, the relativistic correction to the vector meson wave function, which is discussed in Section \ref{sec:LO}, is not expected to be significant \cite{Hood}.

We noted that an alternative probe of the low $x$ gluon at low scales has been
considered in \cite{Blumlein,Gauld}. There they study the data for charm (and
beauty) production obtained in the forward direction by the LHCb
Collaboration \cite{LHCbcc,LHCbbb}. Again it is found that the predictions
strongly depend on the choice of factorization scale. Their prediction
is lower than the data if the natural scale $\mu_F^2=m_Q^2+p_t^2$ is
chosen, and it is found that a larger $\mu_F$ is needed to reproduce the
data using the existing global PDFs. Thus our expectation of a larger
gluon at low $x$ is not in contradiction with the LHCb charm (and beauty)
forward data.

\section*{Acknowledgements}
MGR thanks the IPPP at Durham University for hospitality.  MGR is
supported by the RSCF grant  14-22-00281. SPJ is supported by the
Research Executive Agency (REA) of the European Union under the Grant
Agreement PITN-GA2012316704 (HiggsTools), and TT is supported by STFC 
under the consolidated grant ST/L000431/1.

\end{document}